\documentclass[pre,floatfix,twocolumn]{revtex4} 

\usepackage{epsfig}
\usepackage{graphicx}

\begin{document}

\author{Brian Utter}
\email{utter@phy.duke.edu}
\author{R. P. Behringer}
\affiliation{Department of Physics and Center for Nonlinear  
and Complex Systems, Box 90305, 
Duke University, Durham, NC 27708}
\date{\today}
\title{Self-diffusion in dense granular shear flows}

\renewcommand{\textfraction}{0.02}
\renewcommand{\topfraction}{0.98}
\renewcommand{\bottomfraction}{0.98}
\setcounter{bottomnumber}{4} 
\setcounter{topnumber}{4}
\renewcommand{\floatpagefraction}{0.95}

\newcommand{\ea}{{\it et al.}}

\begin{abstract}		
Diffusivity is a key quantity in describing velocity fluctuations in
granular materials.  These fluctuations are the basis of many
thermodynamic and hydrodynamic models which aim to provide a
statistical description of granular systems.  We present experimental
results on diffusivity in dense, granular shear flows in a 2D Couette
geometry.  We find that self-diffusivities $D$ are proportional to the
local shear rate $\dot{\gamma}$ with diffusivities along the direction
of the mean flow approximately twice as large as those in the
perpendicular direction.  The magnitude of the diffusivity is $D
\approx \dot{\gamma} a^2$ where $a$ is the particle radius.  However,
the gradient in shear rate, coupling to the mean flow, and strong drag
at the moving boundary lead to particle displacements that can appear
subdiffusive or superdiffusive.  In particular, diffusion appears to
be superdiffusive along the mean flow direction due to Taylor
dispersion effects and subdiffusive along the perpendicular direction
due to the gradient in shear rate.  The anisotropic force network
leads to an additional anisotropy in the diffusivity that is a
property of dense systems and has no obvious analog in rapid flows.
Specifically, the diffusivity is supressed along the direction of the
strong force network.  A simple random walk simulation reproduces the
key features of the data, such as the apparent superdiffusive and
subdiffusive behavior arising from the mean velocity field,
confirming the underlying diffusive motion.  The
additional anisotropy is not observed in the simulation since the
strong force network is not included.  Examples of correlated motion,
such as transient vortices, and L\'{e}vy flights are also observed.
Although correlated motion creates velocity fields which are
qualitatively different from collisional Brownian motion and can
introduce non-diffusive effects, on average the system appears simply
diffusive.
\end{abstract}


\maketitle



\section{Introduction} 
\subsection{Overview}

Despite the prevalence of granular materials in nature and industry, a
coherent understanding of granular flows is still lacking.
Particularly in dense systems, features such as jamming, shear bands,
and the coexistence of solid- and liquid-like regions make it
difficult to offer a simple theoretical description.  Fluctuations in
both the force network and particle velocities can be of the same
magnitude as the mean values and are known to be important aspects of
the microscopic behavior of dense granular
flows\cite{Howell.ea:99:Fluctuations}.  Due to the complexity of these
systems, one of the key goals of current research is to develop a
statistical description of steady state behavior, such as a
thermodynamic or hydrodynamic model.
Fundamental to statistical approaches is understanding the mean
fluctuating part of particle motion, which is described by a granular
diffusivity.

Sheared granular systems have received considerable attention recently
\cite{Mueth.ea:00:Signatures,Losert.ea:00:Particle,Howell.ea:99:Fluctuations}
as an important example of granular flow.  Diffusion, in particular,
has been studied in a variety of granular systems, such as vibrated
grains \cite{Barker.ea:93:Transient,Wildman.ea:99.Self}, tumblers
\cite{Seymour.ea:00:Pulsed,Hill.ea:03:Structure}, chute flows
\cite{Hsiau.ea:93:Shear,Menon.ea:97:Diffusing,Natarajan.ea:95:Local},
and sheared systems
\cite{Hsiau.ea:99:Fluctuations,Hsiau.ea:02:Stresses,Hunt.ea:94:Particle,Makse.ea:01:Thermodynamic,Radjai.ea:02:Turbulent,Savage.ea:93:Studies,Garzo.ea:02:Tracer,Campbell:97:Self,Latzel.ea:03:Comparing},
but these studies have predominantly focused on rapid flow regimes.
Understanding slow, high density, flow is not trivial
\cite{Campbell:97:Self}; there is no replacement at the fundamental
level for collisionally based kinetic theories that are expected to
apply only in the dilute rapid flow regime.

In this paper, we characterize the self-diffusivity of grains in a 2D
Couette shearing experiment by studying individual particle
trajectories over time.  In contrast to most previous results, we
focus on quasistatic dense flows.  In this regime, since particles are
constantly in contact with their neighbors, interactions are not
collisional and material flow is largely confined to a shear band with
a nominal thickness on the order of 5 particle diameters.

Several observations from the present experiments are noteworthy: 1)
We find that particle diffusivity is proportional to the local shear
rate, with diffusivities approximately twice as large along the mean
flow direction as the perpendicular direction.  2) We show that unlike
rapid flows, the anisotropic force network induces a substantial
anisotropy in the diffusivity.  This is in addition to the usual
anisotropy induced by the direction of mean flow.  3) Care must be
taken when calculating diffusivities in a shear gradient
\cite{Sierou.ea:03:Shear}, as motion can appear to be subdiffusive or
superdiffusive due to a gradient in the shear rate or Taylor
dispersion \cite{Taylor:53:Dispersion}.  We show through a simple
Fokker-Planck model that apparent sub- or super-diffusive behavior can
be attributed to shear gradient and boundary effects.  4) Hence, the
grain motion is statistically consistent with a simple random walk in
the presence of shear gradients.  5) Nevertheless, at larger spatial
scales, we occasionally observe correlated motion and L\'{e}vy
flights.  But these events are rare and do not have a significant
impact in the mean.

\subsection{Models for Granular Diffusion}
\label{sec-diff-models}

In the kinetic theory approach, a granular temperature is often defined
as T $\propto$ $\langle(v - \overline{v})^2\rangle$, with
instantaneous particle velocity $v$ and mean velocity $\overline{v}$,
in which the velocity fluctuations contribute to a temperature in
analogy with molecular gases.  

A different approach was recently proposed by Makse and Kurchan, who
applied uniform shear in a numerical experiment and measured
diffusivity $D$ and mobility $\chi$ to define a temperature $D/\chi$
by analogy with fluid systems \cite{Makse.ea:01:Thermodynamic}.  In
their model, they report that the $0^{th}$ law (thermal equilibration)
is satisfied in a bidisperse mixture, supporting the thermodynamic
picture.  This is in contrast to experimental measurements of kinetic
granular temperature which find a lack of equipartition when different
types of particles are present
\cite{Feitosa.ea:02:Breakdown,Wildman.ea:02:Coexistence}.  

Isotropic Brownian diffusion in an unbounded system is often
characterized by the time evolution of the second moments of a
probability distribution function (PDF).  For example,
\begin{equation}
\langle x^2\rangle = 2 D t
\end{equation}
where $x$ is the particle position relative to its initial 
position ($v \equiv \Delta
x/\Delta t$ for a small time step $\Delta t$), $D$ is the diffusivity,
and $t$ is time.

More generally, diffusion must be described by a tensor.  For
instance, diffusivities along the flow direction in granular gases are
in general different from transverse diffusivities
\cite{Garzo.ea:02:Tracer,Hsiau.ea:02:Stresses,Campbell:97:Self}.

Diffusion in even a simple shear flow is complicated by Taylor
dispersion effects \cite{Taylor:53:Dispersion}, in which diffusive
motion couples to the mean flow leading to larger dispersion along the
flow direction, as recently elucidated in systems of noncolloidal
particles \cite{Sierou.ea:03:Shear}.  In this case, $\langle
x^2\rangle$ is nonlinear in time, i.e. it contains higher order
corrections due to the coupling of the shear to the diffusive motion.
Simply subtracting the mean flow from particle trajectories and
computing diffusivities does not give accurate results in this system
\cite{Sierou.ea:03:Shear}.

In particular, for flow of the form $\vec{v} = \dot{\gamma} y
\hat{x}$, i.e. uniform unbounded shear flow in two dimensions
in which there is a constant shear rate
$\dot{\gamma}$ creating a velocity gradient in the $y$ direction, the
second order moments are given by
\cite{Elrick:62:Source,Sierou.ea:03:Shear}:
\begin{eqnarray}
\langle yy \rangle &=& 2 D_{yy} t
\label{Dyy-eqn} \\
\langle xy \rangle &=& 2 D_{xy} t + D_{yy} \dot{\gamma} t^2 
\label{Dxy-eqn} \\
\langle xx \rangle &=& 2 D_{xx} t + 2 D_{xy} \dot{\gamma} t^2 + 
\frac{2}{3} D_{yy} \dot{\gamma}^2 t^3 
\label{Dxx-eqn}
\end{eqnarray}
These equations describe an ensemble average of particle positions 
$(x(t), y(t))$ relative to the particle's initial location (i.e.
((x(0), y(0)) = (0,0)), without subtracting the mean flow.

These relations follow naturally for a PDF described by Brownian
anisotropic diffusion with mean local flow $\vec{v}$ as above, and a
diffusion tensor, $\bf{D}$, with elements $D_{xx}$, $D_{yy}$ and
$D_{xy} = D_{yx}$ where 

\begin{equation}
\partial P/\partial t = \vec{v} \cdot \nabla P + \nabla \cdot {\bf D} \nabla
P.
\end{equation}

Here, $x$ corresponds to the streamwise direction and $y$ is
perpendicular 
to $x$.  The left hand sides of Eq.~\ref{Dyy-eqn}-\ref{Dxx-eqn} 
correspond to  
the $y^2$, $xy$, and $x^2$ moments of $P$.

The higher order terms in Eq.~\ref{Dxx-eqn}
are due to Taylor dispersion.  For instance, the 
$t^3$ term arises because diffusive motion along
$\pm \hat{y}$ moves grains to regions of different mean velocity 
$v(y) \hat{x}$ which tends to increase their separation or dispersion 
along the $\hat{x}$ direction.  These higher order terms contribute to 
mean squared displacements which therefore appear superdiffusive.

It must be emphasized that equations~\ref{Dyy-eqn}-\ref{Dxx-eqn} are derived
assuming that the diffusivities and shear rate are constant in 
space and time over
an infinite plane.  This condition is not met in the current
experiment which changes the specific form of the correction terms.

Several issues concerning a diffusive picture must be addressed for
sheared dense granular materials to determine whether Brownian
diffusion applies.  Two of these issues are the presence of a shear
band and the limiting boundary at the shearing surface.  Even assuming
that a diffusive description is applicable, it remains to be
determined on what temporal or spatial scales such a description
should apply.  In dense quasistatic flows, grains are generally close
to a jammed state in which particles are in constant contact.  Motion
of grains requires the creation of voids, so correlated motion might
be expected to be particularly important in dense 2D systems where
paths are constrained.  Short-lived vortex structures have been seen
in 2D granular simulations \cite{Radjai.ea:02:Turbulent} and
experiments on 2D shearing of foams\cite{Debregeas.ea:01:Deformation}.
These are potential deviations fom Brownian diffusive behavior which
might affect the time evolution of the moments.

\subsection{Previous Measurements and Simulations}

The full diffusion tensor has been measured in granular gases using
kinetic theory \cite{Garzo.ea:02:Tracer}, simulations of rapid
granular shear \cite{Campbell:97:Self}, and shearing of non-colloidal
suspensions \cite{Breedveld.ea:02:Measurement}.  

Substantial work on granular diffusivity in rapid flows has been done
by Hsiau and coworkers who have measured self-diffusion coefficients
in a variety of granular systems
\cite{Hsiau.ea:93:Kinetic,Hunt.ea:94:Particle,Hsiau.ea:99:Fluctuations,Hsiau.ea:02:Stresses}.
They find that fluctuations are anisotropic, with the largest
fluctuations along the flow.  Diffusivities are found to increase with
shear rate and depend on the square root of the granular temperature T
in agreement with kinetic gas theory.  
Other results in a similar chute flow were subsequently presented by
Natarayan \ea \cite{Natarajan.ea:95:Local}.  

Losert \ea~studied a 3D fluidized Couette experiment
\cite{Losert.ea:00:Particle}, in which velocity fluctuations were
found to be slightly larger in the direction along the mean flow.
These fluctuations decrease roughly exponentially far from the inner
cylinder, but decrease more slowly than the average velocity.  

Radjai and Roux studied particle velocity fluctuations in numerical
simulations under homogeneous strain in which there was no shear band
formation \cite{Radjai.ea:02:Turbulent}.  They measured anomalous
diffusion with an exponent of 0.9 (rather than 0.5 for ordinary
diffusion) which they attributed to long-time configurational memory
of a granular medium in quasistatic flows.

Diffusivities have also been measured in a 3D rotating tumbler
\cite{Seymour.ea:00:Pulsed,Hill.ea:03:Structure}, 
2D swirling flow \cite{Scherer.ea:96:Swirling},
chute flow \cite{Menon.ea:97:Diffusing}, simulations of shaken spheres
\cite{Barker.ea:93:Transient}, and simulations of small numbers of
spheres in suspension \cite{Marchioro.ea:01:Shear}.  Earlier studies
primarily addressed rapid flows from kinetic theory
\cite{Savage.ea:93:Studies}.

Although these studies are relevant here, we note that the
displacements were assumed to be purely induced by the shear flow and
no attempt was made to investigate the role of the force chain
network.

\subsection{Organization of Presentation}

The paper is organized as follows.  In Section~\ref{sec-experiment},
we describe the experimental techniques.  We present diffusion
measurements in Section~\ref{sec-diffusion} and results from a random
walk simulation in Section~\ref{sec-simulation}.  We show the impact
of the anisotropic force network in Section~\ref{sec-anisotropic}.  We
discuss diffusivities determined from the velocity autocorrelation
functions in Section~\ref{sec-velautocorr}.  In
Section~\ref{sec-correlated}, we show examples of intermittent
vortices and L\'{e}vy flight trajectories, and in
Section~\ref{sec-conclusions}, we draw conclusions.

\section{Experimental Techniques}
\label{sec-experiment}

\begin{figure}
\includegraphics[width=3.1in]{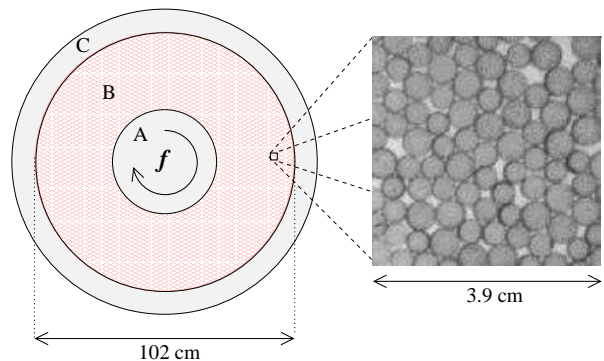}%
\caption{\label{expt-schematic} Schematic of experiment as viewed from
above.  The granular material (B) is contained by the shearing wheel
(A) and the outer ring (C).  Their radii are $R_i$ = 20.5 cm and $R_o$
= 51 cm respectively.  
On the right is a section of an experimental image of the grains.  }
\end{figure}

The experiment is performed with a 2D Couette apparatus, as sketched
in a top view in Fig.~\ref{expt-schematic}.  The granular material (B)
consists of a bidisperse mixture of about 40,000 disks (diameters
$d_S$ = 0.42 cm, $d_L$ = 0.50 cm, thickness = 0.32 cm) in a ratio of 3
small : 1 large.  The bidisperse mixture is used to inhibit
crystalline ordering of the disks.  The disks lie flat on a Plexiglas
sheet bounded by an outer ring ($R_o \equiv$ 51 cm)(C) and an inner shearing
wheel ($R_i \equiv$ 20.5 cm) (A).  A Plexiglas sheet covers the experiment to
protect the experiment from external perturbations, but the sheet does
not contact the particles.  The shearing wheel is rotated at a
frequency $f$ of 0.1-10.0 mHz or a speed of $v \approx 0.013 - 1.3$
cm/s at the shearing surface.  The experiment is initially run for at
least one revolution of the shearing wheel in order to avoid effects
from transients, an issue that will be addressed in another paper
\cite{Utter.ea:03:Transients}.  The shearing wheel and the outer ring
have teeth with gaps comparable to the size of the smaller particles.

The system is lit from below and observed from above using a 2
megapixel CCD camera at a frame rate of up to 7.5 Hz.  Sequences of
approximately 1500 images are analyzed to determine particle
trajectories.  Disks within about 20 particles diameters of the
shearing wheel are marked with lines, thus allowing us to track
particle position and orientation and to identify particles by size.
Approximately 4000 grains are typically in the field of view, of which
2500 are marked.  Images are threshholded and the orientation and
position of lines on the disks are found.  Sufficient temporal
resolution is used such that each particle in a frame can be connected
with the closest grain in the subsequent frame to establish particle
trajectories.

The force network can also be visualized since the grains are made of
a photoelastic material \cite{PSM4}.  When polarized light travels
through the disks, it experiences a phase shift (birefringence)
proportional to the difference in principle stresses $\sigma_2 -
\sigma_1$.  When the disks are illuminated between crossed polarizers,
grains under larger stress are seen as regions of larger gradients in
light intensity.  In this way, the force network is visualized as a
network of bright lines on a dark background.  Additional details were
presented by Howell \ea \cite{Howell.ea:99:Fluctuations}.  The
polarizers are removed for measuring particle trajectories and
diffusivities.

\section{Diffusion Measurements}
\label{sec-diffusion}

\subsection{Mean Velocity Profiles}

\begin{figure}
\includegraphics[width=3.1in]{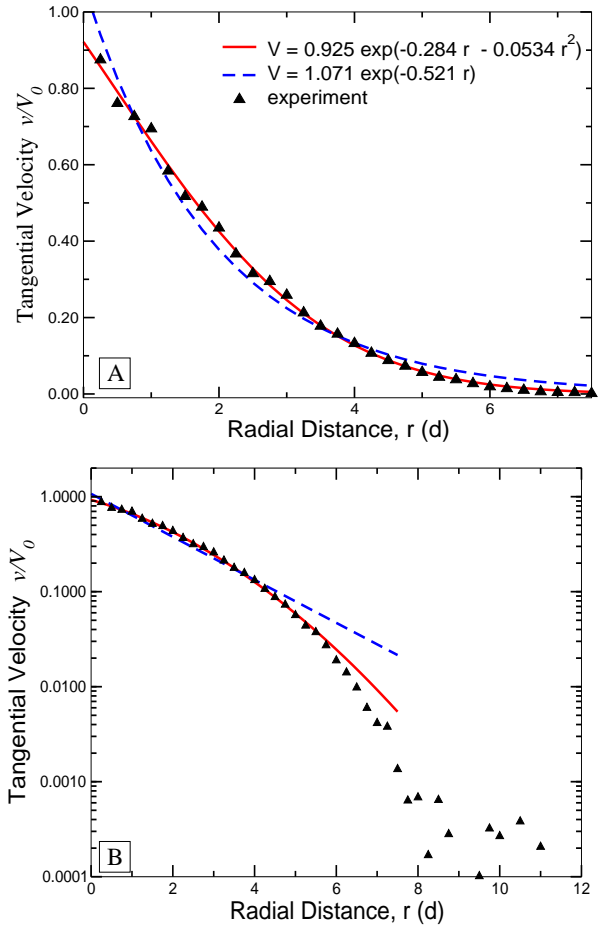}%
\caption{\label{VvsR} Mean tangential velocity vs. radial distance from 
the shearing surface 
plotted on a (A) linear and (B) logarithmic scale for a particular
run.  The velocity $v$ is scaled by the velocity of the shearing surface
$V_0$.
The experimental data ($\triangle$, $r <$ 7.5 d) is fitted to an
exponential (dashed line) and a gaussian (solid line).  $f$ = 1 mHz 
and $V_0 = 0.28 d/$s.
}
\end{figure}

We first consider the mean properties of the flow.  In
Fig.~\ref{VvsR}, we show the mean tangential velocity, $v_\theta$,
versus radial distance from the shearing surface, $r \equiv R - R_i$, 
where $R$ is the distance from the center of the shearing 
wheel.  The velocities are
scaled by the velocity of the shearing surface $V_0 = 0.28 d/$s where
$d$ is the mean particle diameter ($d = (d_S + d_L)/2$).  For this
particular run, we used a frame rate of 1.08 Hz, so that the shearing
wheel was displaced 0.25 $d$ between each of the 1080 images, which,
in total, correspond to one revolution of the shearing wheel.  The
limiting value for $v_\theta$ of approximately $10^{-4}$ corresponds
to the sensitivity of the measurement for a typical number of images
and spatial resolution, e.g. for an image resolution of 20 pixels per
diameter, a grain displacement of 1 pixel over the entire run would
give a mean velocity $\overline{v_\theta} = \frac{1}{1080} \cdot
\frac{d}{20} \cdot 1.08$ Hz $= 5 \times 10^{-5} \frac{d}{s}$ (or
$v/V_0 \approx 10^{-4}$) and velocities smaller than this cannot be
resolved.  Motivated by previous results for Couette shear
\cite{Mueth.ea:00:Signatures,Veje.ea:99:Kinematics}, we fit the data
($r <$ 7.5 d) to an exponential ($v_\theta (r)$ = 1.071
exp(\mbox{-0.521 $r$})) and to a gaussian ($v_\theta (r)$ = 0.925
exp(-0.284 $r$ -~0.0534 $r^2$)) .  

\begin{figure}
\includegraphics[width=3.1in]{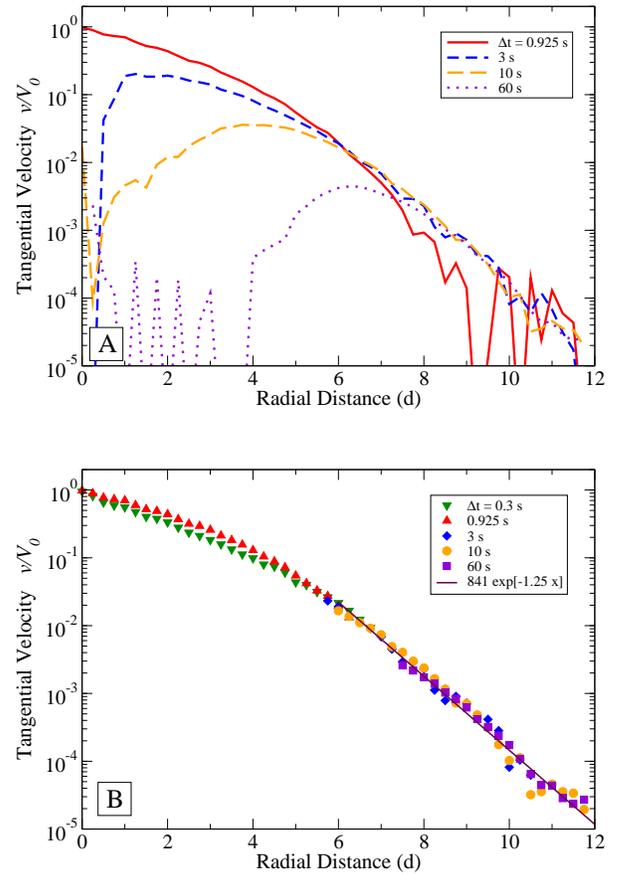}%
\caption{\label{VvsR-1mHzXXs} 
(A) Tangential velocity versus radial
distance from the shearing surface 
using experimental runs at different frame rates ($\equiv
\frac{1}{\Delta t}$) as described in text.  Velocities $v$ are 
shown relative to the the velocity of the shearing surface $V_0 = 0.28$ d/s.
(B) Data is shown where velocities can be resolved given the frame
rate and number of pictures.
An exponential tail is
observed in which $v_\theta \propto e^{-1.25 r}$.
}
\end{figure}

From this individual run, the
Gaussian fit seems most appropriate.  However, 
this is an artifact of the data resolution.
Additional data with slower frame rates points to an important issue
concerning particle tracking velocimetry.  If the number of images $N$
remains fixed, by taking data using slower frame rates, velocities of
slower particles further from the shearing surface can be accurately
measured while faster particles at the shearing surface can no longer
be accurately tracked.  We show these results in
Fig.~\ref{VvsR-1mHzXXs} in which the frame rate ($\equiv 1/\Delta$t)
is varied for different runs at the same imposed shear rate ($f$ = 1
mHz, $N$ = 1500).  The individual curves are accurate over a
particular range of velocities based on the frame rate and number of
pictures in the run.  In Fig.~\ref{VvsR-1mHzXXs}B, we show data for
each set within this range.  It becomes evident that the velocity
profile has an exponential tail which is obscured when simply
analyzing a single run.  Previous results have shown exponential
\cite{Latzel.ea:03:Comparing,Howell.ea:99:Fluctuations},
Gaussian 
\cite{Mueth.ea:00:Signatures}, and similar strongly decaying 
\cite{Losert.ea:00:Particle}
velocity profiles for Couette flow.
Authors of these studies \cite{Mueth.ea:00:Signatures,Howell.ea:99:Fluctuations} 
have suggested that the
differences in measured profiles may depend on whether the flow is 2
or 3D or on whether the particles are rough or smooth.  The present
data suggest that an additional factor may be spatio-temporal resolution.
Particle tracking issues in particular have been addressed  
recently by Xu \ea \cite{Xu.ea:03:Measurement}.

We conclude that correct tracking occurs for velocities that
approximately satisfy
\begin{equation}
\frac{1}{N}\frac{1}{20}\frac{d}{\Delta t} \stackrel{<}{\sim} \overline{v_\theta} 
\stackrel{<}{\sim} 0.1 \frac{d}{\Delta t}
\end{equation}
where $N$ = number of pictures, and the image resolution is 20 pixels
per diameter $d$.  The lower speed limit is set by image and temporal
resolution.  The upper limit is chosen to resolve the occasionally
fast displacements well above the mean.  Note, however, that the upper
cutoff is not an issue when the speed of the shearing surface is less
than $d/\Delta t$, since generally, all particle displacements can be
resolved, e.g. for $\Delta t$ = 0.3 s and 0.925 s, the requirement
that $\overline{v_\theta} < 0.1 \frac{d}{\Delta t}$ is not necessary.

\subsection{Radial and Tangential Diffusivities, $D_{rr}$ and 
$D_{\theta\theta}$}

We measure diffusivities by tracking individual particles, and hence
their displacements, over time in both the radial ($\hat{r}$) and
tangential ($\hat{\theta}$) directions.  We subtract the mean
flow (Fig.~\ref{VvsR}) from the tangential velocity component at 
each time step. 

Although Eqs.~\ref{Dyy-eqn}-\ref{Dxx-eqn} characterize absolute $x$
and $y$ displacements without subtracting the mean flow, they are also
predicated on a velocity profile such that $\langle x(t) \rangle =
\langle y(t) \rangle = 0$ which is true for a uniform shear rate in an
infinite domain, neither of which are true here.  The exponential
velocity profile observed in the experiment invalidates this and
prevents us from directly comparing to
Eqs.~\ref{Dyy-eqn}-\ref{Dxx-eqn}.  Deriving the corresponding moment
evolution equations with the exponential profile of the velocity and
diffusion fields is significantly more complicated and beyond the
scope of this paper.  Instead, we use a random walk model in
Section~\ref{sec-simulation} to model the system.  Nonetheless, the
theory above explains the origin of the Taylor dispersion effects that
we observe.  In order to avoid the effects of a mean displacement due
to the locally varying shear rate, we remove the mean flow.  
The resulting mean
displacement squared is plotted versus time and averaged for different
particles initially within the same radial bin (bin size = $d$ or
$d/2$).  An initially linear evolution indicates ordinary diffusive
behavior with the slope of the line equal to 2$D$.

\begin{figure}
\includegraphics[width=3.1in]{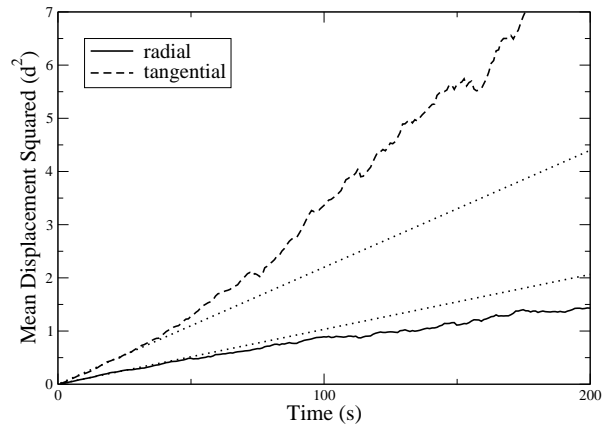}%
\caption{\label{TRdff-l2}  Mean
displacement squared vs. time for tangential and radial directions for
particle trajectories starting at $2d < r < 3d$ and shearing wheel
frequency of $f$ = 1mHz.  Dotted lines show linear fits to t $<$ 30 s.
}
\end{figure}

Fig.~\ref{TRdff-l2} shows a typical example of the mean displacement
squared for the tangential $\langle(R\Delta \theta)^2 \rangle$ and
radial $\langle(\Delta r)^2 \rangle$ directions for particles in the
shear band.  Here, the $\Delta$ notation reminds us that the mean 
flow is subtracted from the data.
The dotted lines are linear fits for $t < 30$s giving
diffusivities proportional to the slopes.  The tangential diffusivity
is approximately double the radial diffusivity at small times.  The
former is expected to deviate from a straight line due to the higher
order terms similar to those 
in Eq.~\ref{Dxx-eqn}.  Note, however, that for small $t$,
the linear term in Eq.~\ref{Dxx-eqn} dominates, and Taylor dispersion 
effects are not present.  

\begin{figure}
\includegraphics[width=3.1in]{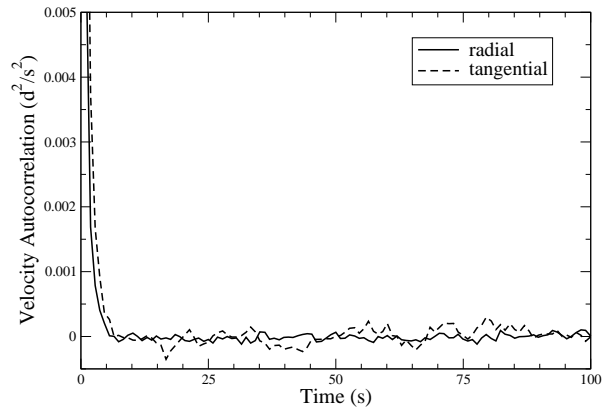}%
\caption{\label{autocorr} Velocity autocorrelation versus time for
particles in $2d < r < 3d$ for radial and tangential velocity
components.
}
\end{figure}

One might worry that using early times would be inaccurate when
diffusivity is generally defined as a long time behavior.  In
particular, results for more rapid flows show an initial ballistic
regime
\cite{Wildman.ea:99.Self,Menon.ea:97:Diffusing,Campbell:97:Self}, and
significant velocity autocorrelations appear in noncolloidal
suspensions \cite{Sierou.ea:03:Shear}.  However, in the quasistatic
motion of the present experiment, there is no ballistic behavior
because grains are constantly in contact with each other.  Moreover,
as seen in the velocity autocorrelation shown in Fig.~\ref{autocorr},
the velocities quickly become uncorrelated.  The time for the
correlation to reach zero corresponds to a mean relative grain
displacement of 0.25 $d$ and occurs within 4 seconds for this data.
Therefore, diffusivities measured for 5s $< t <$ 30s can be expected to
be beyond the correlated regime and are at times before significant
Taylor dispersion effects are observed.  We show below, using a random
walk simulation, that the apparent subdiffusivity of the radial
component in Fig.~\ref{TRdff-l2} is due to the radial gradient in
shear rate, an effect that is not included in
Eqs.~\ref{Dyy-eqn}-\ref{Dxx-eqn} above.

\begin{figure}
\includegraphics[width=3.1in]{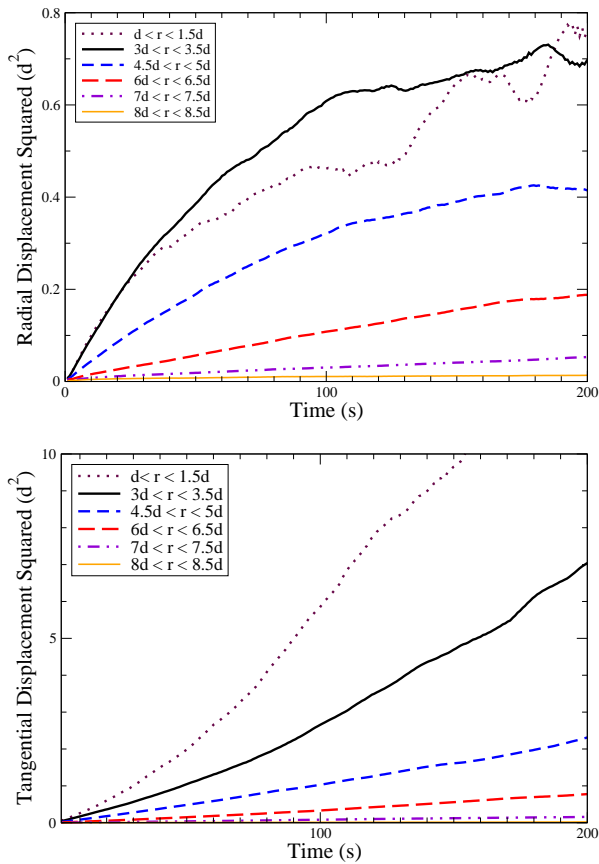}%
\caption{\label{Rdff-alllayers} 
Mean displacement squared vs. time at different distances
from shearing wheel.  Radial displacements shown on top and
tangential displacements on bottom.  $f$ = 1mHz.
}
\end{figure}

\begin{figure}
\includegraphics[width=3.1in]{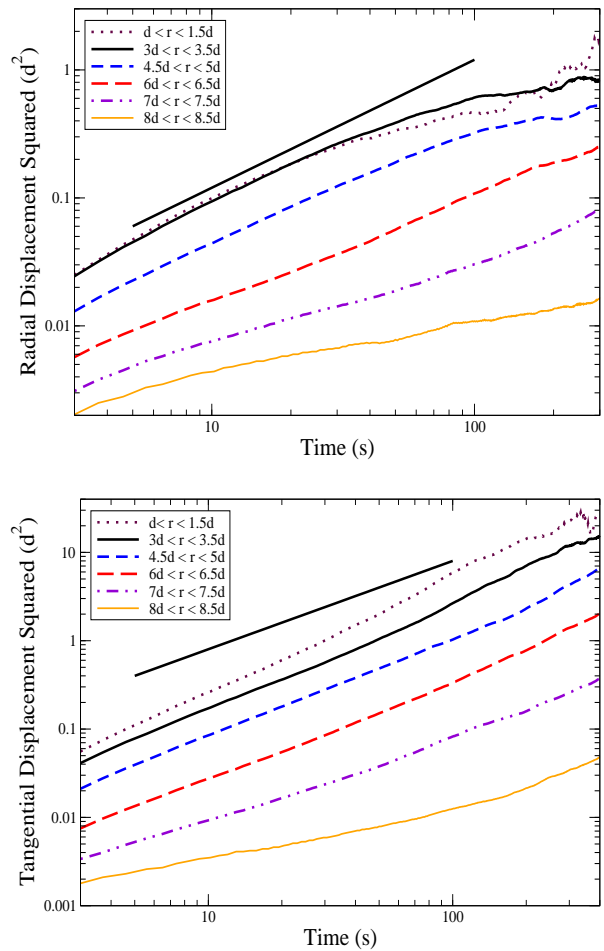}%
\caption{\label{Rdff-all-log} Log-log plot of displacement squared
versus time for data in Fig.~\ref{Rdff-alllayers}.  
Radial displacements given in plot on top and
tangential displacements on bottom. The straight solid line gives the
expected behavior for diffusive motion. \vspace{0.3 in}}
\end{figure}

Fig.~\ref{Rdff-alllayers} shows $\langle(R\Delta \theta)^2 \rangle$
and $\langle(\Delta r)^2 \rangle$ for particles initially at various
distances from the shearing surface.  It is evident that the
diffusivity is larger close to the shearing surface, i.e. in regions
of large shear, as observed previously.  In addition, the maximum
diffusivity occurs at $r \approx 2d$.  The fact that the maximum
diffusivity does not occur at $r = 0$ is due to non-diffusive motion
of particles in contact with the shearing wheel.  Since most of these
particles ($r < 2d$) are dragged by the wheel at the same speed and
the mean velocity has been subtracted, the fluctuations are smaller.
In Fig.~\ref{Rdff-all-log}, we show the RMS displacements versus time
on a log-log scale.  The solid line shows the expected slope for
diffusive behavior.

\vspace{0.2in}

\begin{figure}
\includegraphics[width=3.1in]{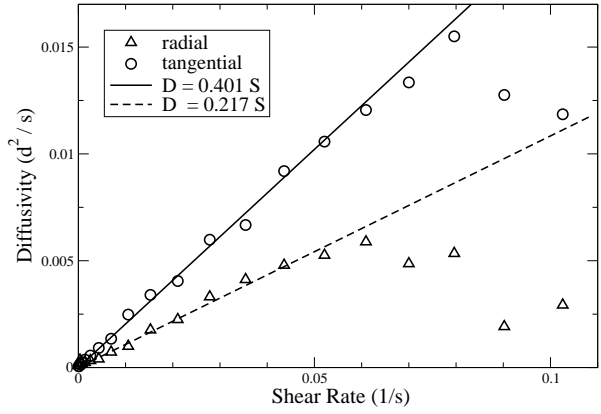}%
\caption{\label{DvsShearRate} 
Diffusivities $D$ vs. local shear rate $\dot{\gamma}$ ($f$ = 1 mHz). 
Diffusivities are proportional to the local shear rate.
Close to the shearing wheel, 
at larger shear rates, the diffusivity 
decreases because particles at $r<2$ are typically 
dragged continuously by the shearing wheel.
The solid line shows $D = 0.200~\dot{\gamma}$ and the dashed line shows 
$D = 0.108~\dot{\gamma}$.
}
\end{figure}

As seen in Fig.~\ref{DvsShearRate}, the diffusivity is proportional to
the local shear rate $\dot{\gamma}$.  For this figure, we use the
local shear rate determined from the slope of Fig.~\ref{VvsR}, and a
diffusivity that is half the slope of $\Delta r^2$ vs. $t$ for $t <$
30 s.  The decrease in diffusivity at large shearing rate (i.e. close
to the shearing surface) is due to particles being dragged by the
shearing wheel and hence exhibiting ballistic behavior.  For the
radial and tangential directions, $D \approx 0.1-0.2~d^2 \dot{\gamma}
\approx 0.4-0.8 a^2 \dot{\gamma}$ where $a$ is the particle radius,
i.e. the scale of the diffusivity is approximately given by $a^2
\dot{\gamma}$.  As a consequence of the exponential tail of the
velocity profile, the diffusivity also decays roughly exponentially,
such that the diffusive motion is effectively confined to the shear
band.

\begin{figure}
\includegraphics[width=3.1in]{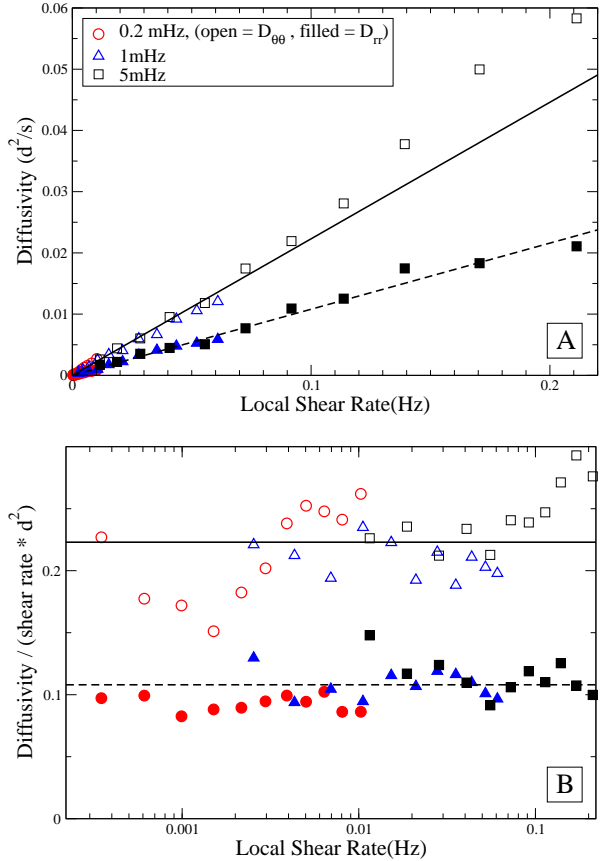}%
\caption{\label{DvsShear-all3-both} 
(A) Diffusivities vs. local shear rate for grains at $r>2$ for three
different rotation rates of the shearing wheel.
In (B), the diffusivities are rescaled by the local shear rate
$\dot{\gamma}$ and $d^2$.  The lines show fits for $D_{\theta\theta}
= 0.223 \dot{\gamma}$ (solid) and $D_{rr} = 0.108 \dot{\gamma}$
(dashed) which approximately hold for three orders of magnitude of 
shear rate.  
}
\end{figure}

Fig.~\ref{DvsShear-all3-both}A shows results for diffusivity versus
local shear rate ($r > 2d$) for three different rotation frequencies
of the shearing wheel.  The diffusivity is approximately proportional
to local shear rate over a large range of shearing rates from separate
experimental runs, and $D_{\theta\theta}/D_{rr} \approx 2$.  In
Fig.~\ref{DvsShear-all3-both}B, the diffusivity at each data point was
divided by the local shear rate.  The resulting data
($D/\dot{\gamma}$) is roughly constant over 3 orders of magnitude of
shear rate.  The lines show fits for tangential diffusivities
$D_{\theta\theta}$ = 0.223 $\dot{\gamma}$ and radial diffusivities
$D_{rr}$ = 0.108 $\dot{\gamma}$.  The data for $D_{\theta\theta}$ is
noisier than that for $D_{rr}$ since due to the mean flow 
the magnitude of the tangential
motion is much larger than radial motion.

\subsection{Off-diaganol Diffusivity $D_{r\theta}$}

\begin{figure}
\includegraphics[width=3.1in]{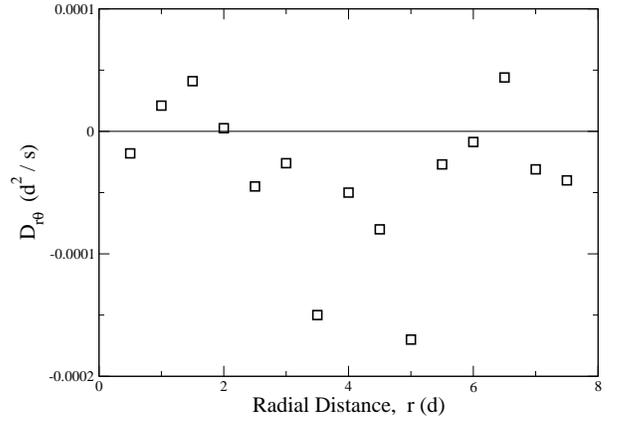}%
\caption{\label{Drt} 
Off-diagonal diffusivity $D_{r\theta}$.
}
\end{figure}

The off-diagonal diffusion constant $D_{r\theta}$ is shown in
Fig.~\ref{Drt}.  This diffusion coefficient is an order of magnitude
smaller than $D_{rr}$ and $D_{\theta \theta}$.  Away from the
shearing surface, $D_{r\theta}$ is also negative.  This is due to the
anisotropic force network and will be addressed below.

\section{Random Walk Simulation}
\label{sec-simulation}

As mentioned earlier, direct comparison of the data to
Eq.~\ref{Dyy-eqn}-\ref{Dxx-eqn} is not possible.  
The fact that $v(r)$ and $D(r)$ decay exponentially and the presence of the 
boundary at $r = 0$ in the experiment are inconsistent with the
assumptions leading to the moment evolution equations
in Section~\ref{sec-diff-models}.
An exact solution of the moment equations
with the appropriate boundary conditions and spatial dependence of
$D(r)$ and $\vec{v}(r)$ is difficult.  However, a numerical simulation with
appropriate spatial dependence in $D$ and $\vec{v}$ using a random
walk model is relatively simple.

In this section, we present such a simulation in which we
assume diffusive motion and impose an exponential velocity
profile and impenetrable 
inner boundary so as to parallel the experiment.  To model
radial diffusion, a walker makes a step each time $\tau$ with equal
probability along $\pm \hat{r}$ with a radial step length L$_r$($r$)
proportional to $(\dot{\gamma} (r))^{1/2}$, where $\dot{\gamma}$ is
the experimentally measured shear rate.  That is, L$_r$($r$) = $c_1
(\dot{\gamma} (r))^{1/2}$, where $c_1$ is a constant.  This imposes a
diffusivity D $\propto$ L($r)^2$/$\tau$ $\propto$ $\dot{\gamma}$($r$)
in agreement with Fig.~\ref{DvsShear-all3-both}.  Radial motion is
bounded by the shearing wheel, so any step that would move a particle
through that boundary 
is automatically forced to be a step away from the shearing
wheel (i.e. towards positive $r$).  Tangential motion is modeled in a
similar way.  At each time step, the walker is advected at the
experimentally measured mean velocity based on its radial position
(Fig.~\ref{VvsR}) and also randomly takes an additional step along
$\pm \hat{\theta}$ with tangential step length L$_\theta$($r$) = $c_2
\dot{\gamma}$($r$)$^{1/2}$ This model contains two free parameters,
$c_1$ and $c_2$, corresponding to the magnitude of $D_{rr}$ and
$D_{\theta\theta}$.  Equivalently, the fit parameters can be thought
of as determining an overall scale factor for the data and the ratio
of radial to tangential diffusivities.  Here, we do not consider
additional anisotropies associated with the force chain network.

\begin{figure}
\includegraphics[width=3.1in]{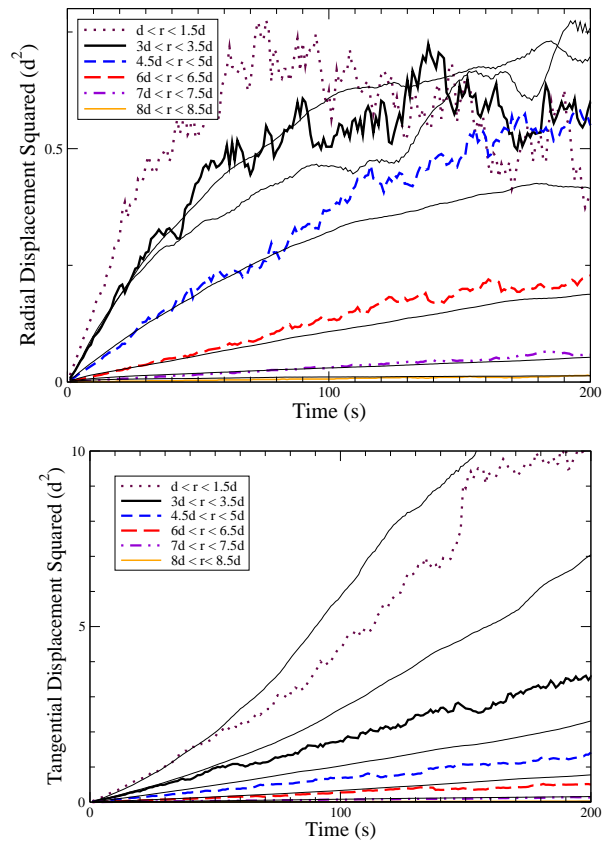}%
\caption{\label{simdata} Mean displacement squared vs. time for
simulation of random walk 
in which the experimentally measured velocity profile is imposed.
Radial displacements
shown on top and tangential on bottom. 
The thin solid lines show the experimental data from 
Fig.~\ref{Rdff-alllayers}.  }
\end{figure}

Fig.~\ref{simdata} shows a mean-square displacement versus time for
the simulated data.  The experimental data from
Fig.~\ref{Rdff-alllayers} is also included for reference as thin solid
lines.  The simulation is performed assuming $D_\theta / D_r = 1.9$
($c_1$ = 0.48, $c_2$ = 0.66).  As noted, the two free parameters set
the scale of the radial and tangential displacements.  We emphasize
that the relative magnitudes of the data at different distances from
the shearing wheel and the apparent subdiffusive and superdiffusive
behavior at longer times result from the experimentally measured
velocity profile.

Although the initial slope of the lines in Fig.~\ref{simdata} 
is equal to 2$D(r)$, the 
long time behavior and deviation from a straight line is due to the 
coupling to the mean flow.  The horizontal and vertical scales of Figures 
\ref{Rdff-alllayers} and \ref{simdata} are identical in order to 
compare the long time behavior.  
This confirms that the apparent subdiffusion and superdiffusion 
is due to the mean flow.

In particular, it is clear that the curvature of the $D_{rr}$ data for
particles close to the shearing surface, which appeared to be
subdiffusive, arises from the gradient in local shear rate.  That is,
the grains next to the wall diffuse away to a region of slower
shearing rate and, once away from the wall, diffuse more slowly.  If
this gradient is removed from the simulation (i.e. $\dot{\gamma}$ is
assumed constant), the lines become straight with approximately the
same slope.  The presence of a wall ($r$ = 0) also tends to decrease
the diffusivity at small $r$, but this is a much less pronounced
effect than that of the gradient in shear rate.  Note 
in Fig.~\ref{simdata} that simulated
grains close to the shearing wheel show an apparent superdiffusive
behavior due to Taylor dispersion.

We note that in the experiment, grains at $r<2$ are generally dragged
by the shearing wheel which tends to decrease the apparent radial
diffusivity and add to the effects of Taylor dispersion (which
accounts for the slight difference in the magnitudes of
Fig.~\ref{Rdff-alllayers} and~\ref{simdata}).  However,
Fig.~\ref{simdata} reveals that the dominant effects are the shear
gradient and Taylor dispersion.  Thus, the main features of the
apparent subdiffusive and superdiffusive behavior are observed even
though the simulation does not include the effect of ballistic motion
due to dragging of particles by the shearing wheel.

\begin{figure}
\includegraphics[width=3.1in]{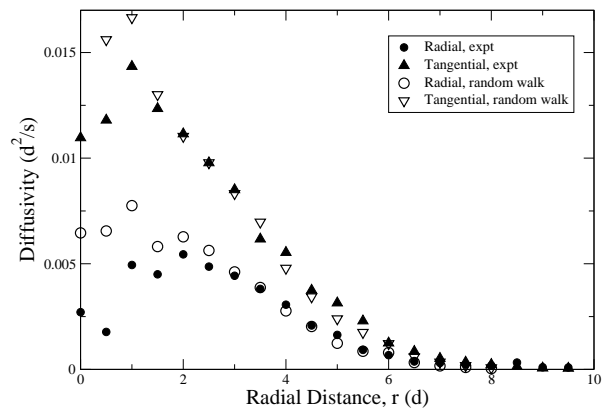}%
\caption{\label{DvsR-all} Diffusivity vs. radial distance from the 
shearing surface for
experimental data and the random walker model.  The deviation between
simulation and experiment for $r<2$ is due to particles being dragged
nondiffusively by the shearing wheel.  }
\end{figure}

We show the measured diffusivities for the experiment and the random
walk simulation together in Fig.~\ref{DvsR-all}.  There is very good
agreement except for $r<2$, where the simulation overestimates the
diffusivity.

\section{Effects Due to Anisotropic Force Network}
\label{sec-anisotropic}

\begin{figure}
\includegraphics[width=3.3in]{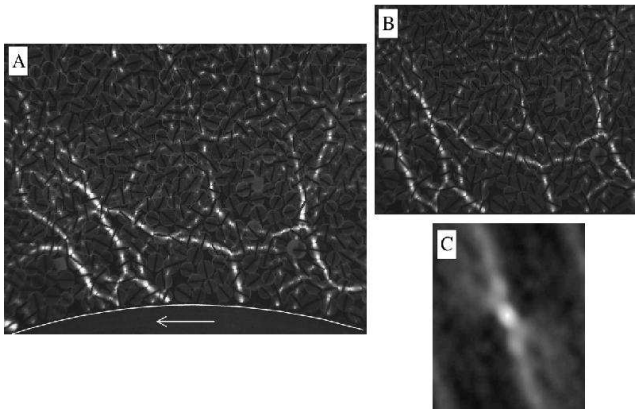}%
\caption{\label{stress-chains-auto} 
(A) Force chains imaged using photoelastic
grains. The shearing wheel is marked by the white line and is rotating
to the left.
(B) Image (A) is rescaled from polar to cartesian coordinates such that 
the shearing wheel is at $y=0$.
(C) A 2D autocorrelation of (B) 
which characterizes the orientation of the 
strong force network.  
}
\end{figure}

In the previous discussion, we tacitly assumed that the natural
coordinate sytem for diffusion measurements is set by the radial and
tangential directions, corresponding to the anisotropy of the imposed
shear.  However, dense systems, unlike dilute rapid flows, have
anisotropic force networks due to imposed shear which are in general
at a different orientation from the flow direction.  This is seen by
using the photoelasticity of the grains to image the force chains, as
in Fig.~\ref{stress-chains-auto}A.  This figure shows a typical case
where the force chains are preferentially oriented to oppose the
motion of the shearing wheel, at an angle that is intermediate between
the $\hat{r}$ and $\hat{\theta}$ directions.  We might expect the
minimum diffusivity to be oriented along the mean direction of the
force chains.  To determine this angle, images such as
Fig.~\ref{stress-chains-auto}A can be transformed from polar to
cartesian coordinates, such that the shearing wheel is located at y =
0 (Fig.~\ref{stress-chains-auto}B).  In this figure, each vertical
line corresponds to a radial line in the original image.  Since the
curvature of the wheel is relatively small, the transformation is not
dramatic and distortion of the image is small.  A 2D autocorrelation
of image \ref{stress-chains-auto}B, then
provides a measure of the mean force chain orientation 
(\ref{stress-chains-auto}C).  
The mean angle 
$\phi$ of
the force chains fluctuates strongly in time around 
a mean value of $20-30^\circ$
relative to $\hat{r}$.

\begin{figure}
\includegraphics[width=3.1in]{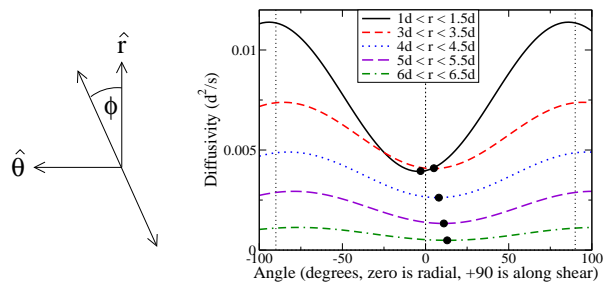}%
\caption{\label{diffvsangle} We project displacements at each time onto
an axis rotated at an arbitrary angle $\phi$ degrees from the radial
direction.  $D_{\phi\phi}$ is measured versus $\phi$ and plotted with
the minimum diffusivity indicated by the filled circles.  
\vspace{0.2 in}}
\end{figure}

To determine the angular dependence of the diffusive motion, we
locally project displacements at each time step onto an axis rotated
at an angle $\phi$ from the radial direction, as sketched in
Fig.~\ref{diffvsangle}.  That is, for each step, the displacement is
locally parameterized in terms of radial and tangential
components (relative to the center of the shearing
wheel) and then are locally projected onto an axis rotated by $\phi$ relative
to the $r$ direction.  We then use the $\phi$ components of the trajectories 
to measure a diffusivity, $D_{\phi\phi}$, along this
direction.  On the right side of Fig.~\ref{diffvsangle}, we show
$D_{\phi\phi}$ versus $\phi$ for grains at different radial distance
$r$.  Again, the diffusivities decrease with distance from the
shearing wheel.  In addition, the direction of minimum diffusivity
$\phi_{min}$, marked by the solid circles, changes with distance from
the shearing wheel.

\begin{figure}
\includegraphics[width=3.1in]{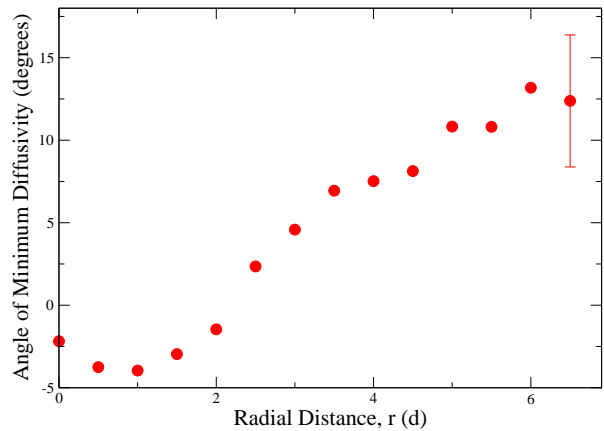}%
\caption{\label{dffangVSr} The angle $\phi$ corresponding to the
minimum diffusivity $\phi_{min}$ is measured from
Fig.~\ref{diffvsangle}.  The increase with radial distance corresponds
to the minimum diffusivity becoming more aligned with the mean direction
of force chains.  }
\end{figure}

In Fig.~\ref{dffangVSr}, we show the angle of minimum diffusivity
$\phi_{min}$ versus distance from the shearing wheel which we
determine by fitting a parabola to $D_{\phi \phi}$ in the region
$\phi_{min} \pm 30^\circ$.  Close to the shearing surface, at high
shear rates, the minimum diffusivity is in the radial direction,
corresponding to the minimum expected based on the imposed shear
direction.  At larger distances $r$, the minimum shifts towards the
direction of the mean force chain orientation.  In other words,
outside of the immediate vicinity of the shearing wheel, the
anisotropic force network affects particle motion and must be taken
into account in order to properly describe the diffusive motion in
dense granular systems.

\begin{figure}
\includegraphics[width=3.1in]{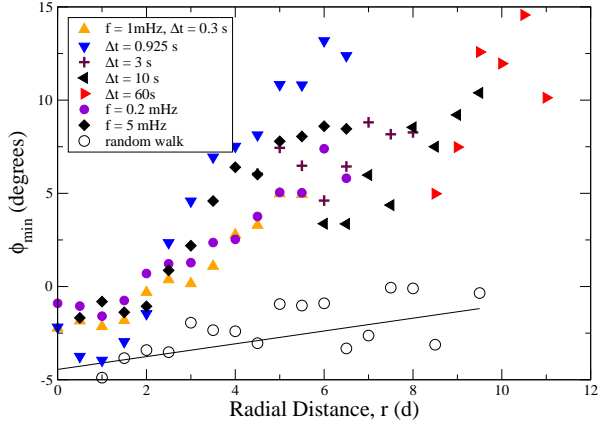}%
\caption{\label{dffangVSr-all} The angle of minimum diffusivity
$\phi_{min}$ vs. $r$ measured according to Fig.~\ref{diffvsangle} for
multiple data sets.  Again, the orientation of minimum diffusivity
shifts towards a direction that corresponds to the mean force chain
direction.  The same analysis is performed on the random walk data 
($\circ$ and linear fit) which does not model the anisotropic force 
network.  
}
\end{figure}

We use the same procedure for deducing the angular dependence of the
diffusivity on a number of independent data sets, including the data
used to create the velocity profile of (Fig.~\ref{VvsR-1mHzXXs}) and
on two additional data sets with different shearing rates.  We show
data in Fig.~\ref{dffangVSr-all} only for velocities that are properly
resolved, as in Fig.~\ref{VvsR-1mHzXXs}.  Although there is some
significant variability from one data set to the next, there is a
clear trend in which $\phi_{min}$ shifts towards the direction of
force chain orientation as $r$ increases.  When we perform the same
analysis on the simulated data, in which there is no force network,
$\phi_{min}$ does not increase above $0^\circ$ as indicated by the
open circles.  The fact that these points are negative is addressed in
the following section.

\section{Diffusivity and velocity autocorrelation functions}
\label{sec-velautocorr}

\begin{figure}
\includegraphics[width=3.1in]{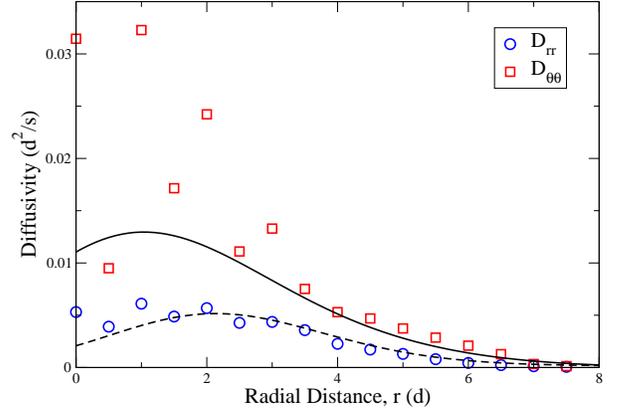}%
\caption{\label{dff-crr} Diffusivities determined from velocity
autocorrelations using Eq.~\ref{autocorr-eqn}.  The shearing wheel
frequency is $f$ = 1 mHz.  The lines show fits to the tangential 
(solid) and radial (dashed) diffusivities in Fig.~\ref{DvsR-all} 
for comparison.
\vspace{.3 in}}
\end{figure}

Diffusivities ($D_{rr}$ and $D_{\theta \theta}$) can also be
determined using velocity autocorrelations
from the expression
\begin{equation}
\label{autocorr-eqn}
D_{xx} = \int_{0}^{\infty} \langle v_x(t) v_x(t + \tau) \rangle d\tau,
\end{equation}
where the velocity at each time step is simply defined as $v_x(t)
\equiv (x(t) -x(t-1))/\Delta t$. The integral must be taken over times
long enough to extend beyond the initial correlated region.  Thus, for
this data,
\begin{equation}
D_{xx} = \Delta t \sum_{dt = 0}^{N} \langle v_x(t) v_x (t + dt) \rangle ,
\end{equation}
where we use a cumulative sum of the autocorrelation
(Fig.~\ref{autocorr}).  After an initial transient of about 5 s, the
curve fluctuates around a constant value.  The average of the data
well after the transient (85 - 185 s, for this data set) is taken as
the diffusivity.  Fig.~\ref{dff-crr} shows the diffusivities
determined from the velocity correlations.  For the most part, they
agree quite well with diffusivities determined from the displacement
squared versus time data which are indicated by lines showing fits to
the data of Fig.~\ref{DvsR-all}.  The exception is for tangential
diffusivities of particles adjacent to the shearing wheel. This
discrepency is due to the fact that 
that there is some non-negligible correlation in
the tangential velocity after $t = 30$s for particles in contact with
the shearing wheel.

\begin{figure}
\includegraphics[width=3.1in]{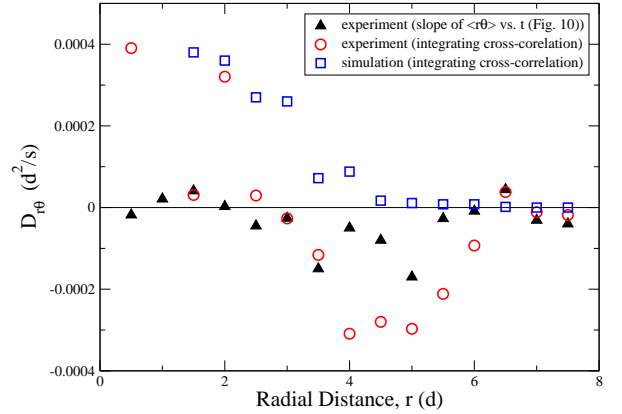}%
\caption{\label{dffRT-crr} Off-diagonal diffusivity determined by
integrating the velocity cross-correlation.  
Data determined from the slope of displacement squared versus time plots 
(Fig.~\ref{Drt}) is shown for comparison.  $D_{r \theta}$ for the random walk
simulation is also shown.  }
\end{figure}

The off-diagonal diffusivity must be determined using
\cite{Batchelor:49:Diffusion}
\begin{equation}
D_{ij} = \frac{1}{2}  \int_0^t (\langle v_i(t') v_j(t) \rangle
+ \langle v_i(t) v_j(t') \rangle) dt'
\end{equation}
which for large $t$ can be rewritten as
\cite{Campbell:97:Self}
\begin{equation}
D_{ij} = \frac{1}{2} \int_{-\infty}^{\infty} \langle v_i(t) v_j(t + \tau) 
\rangle d\tau,
\end{equation}
assuming the motion is statistically stationary over time.  This
reduces to Eq.~\ref{autocorr-eqn} for the diagonal terms, $i = j$.
The off-diagonal diffusivity, $D_{r \theta}$ is shown in
Fig.~\ref{dffRT-crr} along with data determined using the
displacement squared versus time plots (Fig.~\ref{Drt}).  Since the
magnitude of $D_{r \theta}$ 
is small, the fluctuations lead to larger noise in these
results.

Further from the wheel, where the minimum diffusivity shifts to larger
angles, the cross-correlation term is negative.  This indicates that
motion along $\hat{\theta}$ is anticorrelated with motion along
$\hat{r}$, which agrees with a decrease in diffusivity along positive
$\phi$, and a shift in minimum diffusion angle towards positive $\phi$.
To emphasize the effect of the anisotropic force network, we contrast
the experimental results with data from the simulation, where there is
no force network effect and $D_{r\theta}$ is always positive
(Fig.~\ref{dffRT-crr}).

The fact that $D_{xy}$ is positive in the simulation is due to the 
velocity gradient, an effect that was also observed in previous
measurements of the cross-term
\cite{Garzo.ea:02:Tracer,Campbell:97:Self,Breedveld.ea:02:Measurement}.
Note that a positive $D_{xy}$ in the present data corresponds to a
negative $D_{xy}$ in previous results.  This difference in sign is due to
the fact that the authors of
\cite{Garzo.ea:02:Tracer,Campbell:97:Self,Breedveld.ea:02:Measurement}
used the convention that $v_x$ increases with $y$.  By contrast, we
have chosen the natural experimental coordinate sytem $y \equiv r$
such that $y=0$ at the boundary (shearing wheel) and increases into
the bulk, i.e. $v_x$ decreases with $y$.

Returning to the present studies, the fact that $D_{xy}$ is positive
for the simulation is the origin of the slightly negative $\phi_{min}$
seen in Fig.~\ref{dffangVSr-all}.  We emphasize that the positive 
$\phi_{min}$ and negative $D_{xy}$ in the experimental data have the
opposite sign from the simulation.  This difference in sign is due to
the presence of the strong force network in the experiments, an effect
that is absent in the simulations.

\section{L\'{e}vy Flights and Vortices}
\label{sec-correlated}

We also observe examples of correlated motion and trajectories similar
to L\'{e}vy flights \cite{Klafter.ea:96:Beyond} which could contribute
to non-diffusive motion.  In fact, the dense 2D packing leads to
caging and coordinated motion such that neighboring grains tend to
move together.  This differs from more dilated flows in which
collisions are the source of fluctuating motion.  The fact that the
present system behaves diffusively on average indicates that
long-range correlated motion is sufficiently rare and random over time
that the mean behavior is not affected.

It is interesting to ask whether such novel behavior as L\'{e}vy
flights occur in our system and whether they are important.  L\'{e}vy
flights are random walks in which occasional large steps, or flights,
are observed, such that apparent Brownian motion on smaller scales is
punctuated by large displacements.  They also have the property that
the variance of the step size $\langle L^2 \rangle$ and therefore the
diffusivity ($\propto L^2/\tau$) are infinite.  This situation can be
realized if the probability of the walker making a step $L$ is given
by a power-law $P(L) \propto L^{-\alpha}$ where $2 < \alpha < 3$.
This is in contrast with gaussian or exponential distributions of $L$
in which case large steps are much more rare and the variance is
finite.

\begin{figure}
\includegraphics[width=3.1in]{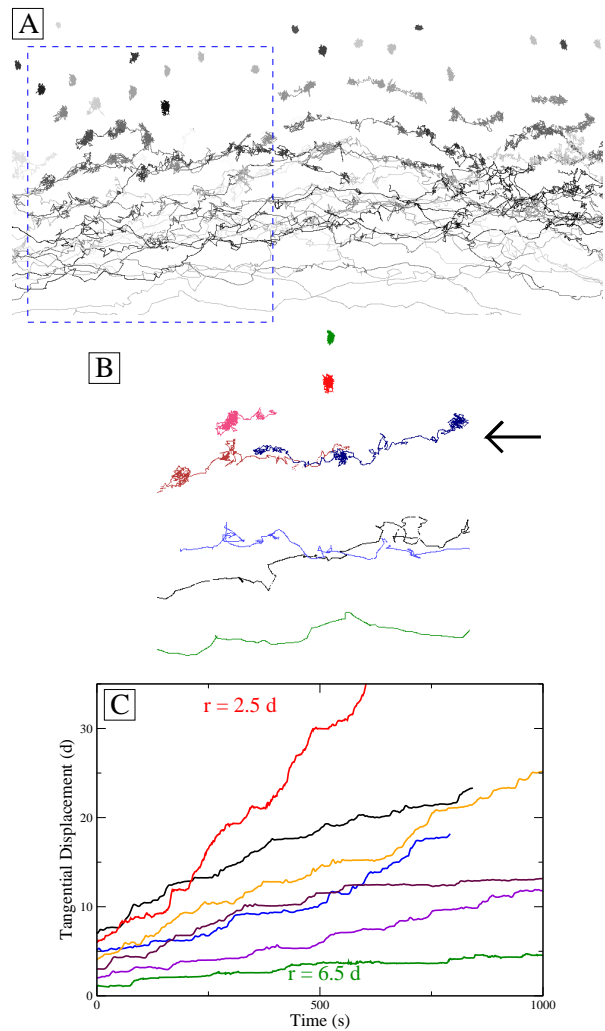}%
\caption{\label{levy} 
(A) Trajectories in the shear band are shown for a fixed time (1000 s).  
(B) A few individual trajectories from the dashed box in (A) are shown.
Motion 
similar to L\'{e}vy flights is occasionally observed ($r \approx 4-5 d$), 
in the region indicated by the arrow. 
Small fluctuations of particle position are observed with occasional 
larger scale advection. 
(C) Tangential displacements versus time for 7 individual trajectories 
($r \approx 2.5, 3, 3.5, 4, 4.5, 5.5, 6.5 d$) chosen to highlight 
L'{e}vy-like motion.
Trajectories closer 
to the shearing wheel are displaced upwards at $t = 0$ for clarity.
}
\end{figure}

In Fig.~\ref{levy}A, we show trajectories for particles in the shear
band.  
Each line shows the trajectory of a single particle over the same 
time period (1000 s).  Next to the shearing surface (bottom of image),
particles travel relatively fast compared to particles outside of the shear 
band (top of image), which fluctuate around effectively stationary 
positions.  
Fig.~\ref{levy}B shows a few particular trajectories from the 
dashed region of Fig.~\ref{levy}A. 
At the edge of this band ($r \approx 4-5 d$), we see
trajectories that are reminiscent of L\'{e}vy flights, in which
relatively large displacements occur between periods of fluctuating
motion on a smaller scale.  To observe this better, in
Fig.~\ref{levy}C, we plot tangential displacement versus time for
seven trajectories at different distances from the shearing wheel.
The data has been smoothed over a 5 s running window.  We note that
these are not necessarily typical trajectories, but have been chosen
to elucidate the presence of L\'{e}vy-like behavior.  In this plot,
regions of small fluctuations (nearly flat lines) are separated by
faster jumps along the mean flow direction.

These data can be compared with an example from a rotating flow fluid
experiment \cite[e.g. Fig.~7]{Solomon.ea:94:Chaotic} from which
similar data were obtained.  Although we observe somewhat similar
motion, the trajectories in Fig.~\ref{levy} have a different origin.
In the fluid case, particles exhibit flights between periods in which
they are trapped by vortices.  In the granular system, grains become
trapped as they move farther from the shearing surface and remain
effectively trapped until they move closer to the wheel.  In addition,
it is common for local rearrangements involving 10-20 grains to occur
intermittently.  (A similar effect may also account for the L\'{e}vy
distributions of trapping times for observations of grains deposited
on sand piles \cite{levy-sandpile}.)  With the present data,
e.g. Fig.~\ref{levy}, we do not have sufficient statistics to
determine whether the trajectories exhibit L\'{e}vy scaling because
flight-like trajectories are rare.

\begin{figure}
\includegraphics[width=3.1in]{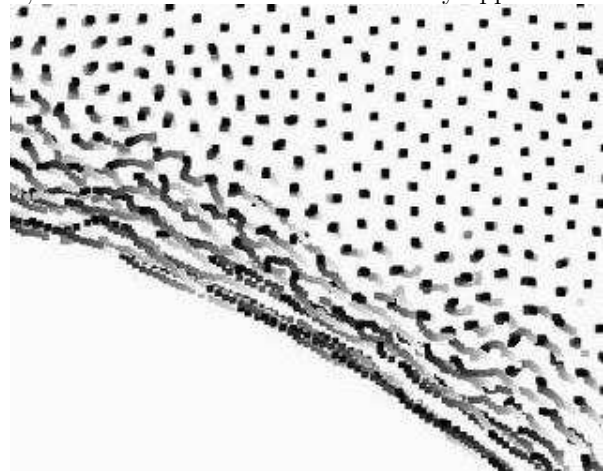}%
\caption{\label{vortex} 
Granular vortices are occasionally observed in plots of particle trajectories
over 25 s.  Particle position is indicated by a dot with darker grayscale 
levels used for later times. 
}
\end{figure}

We also occasionally observe cooperative motion as in
Fig.~\ref{vortex}, which displays particle trajectories over a 25 s
window in which the greyscale level indicates time (light grey = early time,
dark = later time).  In the lower right, there is a region of
locally correlated motion.  A transient vortex is present in the upper
left.  Although correlated motion is common, since motion in a dense
packing requires motion of neighboring grains, vortices are rare
events. In addition, unlike vortices in fluids, there are no inertial
effects and granular vortices appear to quickly dissipate without
affecting the long time behavior of the grains.

\section{Conclusions}
\label{sec-conclusions}

To conclude, we find that granular motion in dense shear flows is
diffusive with a self-diffusivity proportional to the local shear rate
($D \approx \dot{\gamma} a^2$ where $a$ is the particle radius).  However,
the diffusion tensor ${\bf D}$ is anisotropic due to underlying
anisotropies in both the velocity field and force network.  The
velocity anisotropy leads to a tangential diffusivity that is about
double the radial diffusivity.  The anisotropic force network
dominates the local diffusivity outside the immediate vicinity of the
shearing surface, and leads to a minimum diffusivity approximately
along the mean force chain direction.  This latter feature has not
been observed in more rapid flows, to our knowledge, and is a property
of dense granular systems.

Motion can appear subdiffusive due to the decreasing shear rate away
from the shearing surface or superdiffusive due to Taylor dispersion
effects.  A simple random walk model which reproduces the apparent
anomalous diffusion indicates that the underlying motion is diffusive.
Using the experimentally measured velocity profile, assuming
$D_\theta/D_r \approx 2$, and choosing an overall multiplicative scale
factor, the simulation closely matches the experiment, including the
long time behavior which is affected by the gradient in shear rate and
Taylor dispersion.  The simulation also highlights the effects of the
anisotropic force network.  Differences in the sign of $D_{xy}$
between simulation and experiment are associated with the anisotropic
force network.  The same is true for the orientation of the minimum 
diffusivity $D_{\phi\phi}$.  
Velocity autocorrelation plots show that motion in dense
granular flows quickly becomes uncorrelated and there is not a
distinguishable ballistic regime before diffusive behavior dominates.

Examples of correlated motion, such as vortices, and trajectories
similar to L\'{e}vy flights are also observed.  However, these effects
are sufficiently intermittent and random that the system behaves
diffusively.

{\bf Acknowledgement} We appreciate a number of very helpful
discussions with Prof. John Brady.  This work has been supported by
the National Science Foundation through grants DMR-0137119,
DMS-0204677 and DMS-0244492, and by NASA grant NAG3-2372.





%
%

%
%

\end{document}